%% file: paper.tex
\documentclass[conference]{IEEEtran}
\IEEEoverridecommandlockouts
\usepackage{cite}
\usepackage{amsmath,amssymb,amsfonts}
\usepackage{algorithmic}
\usepackage{graphicx}
\usepackage{textcomp}
\usepackage{xcolor}
\def\BibTeX{{\rm B\kern-.05em{\sc i\kern-.025em b}\kern-.08em
    T\kern-.1667em\lower.7ex\hbox{E}\kern-.125emX}}

\usepackage{times}
\usepackage{graphicx}
\usepackage{subfigure}
\graphicspath{{figs/}}
\usepackage{epstopdf}		
\usepackage{longtable}
\usepackage{pdfpages}
\usepackage{bbding}
\usepackage{pifont}
\usepackage{wasysym}
\usepackage{amsmath}
\usepackage{amssymb}
\usepackage{listings}
\usepackage{url}
\usepackage{color}
\usepackage{setspace}
\usepackage{float}
\usepackage{multirow}
\usepackage[]{algorithm2e}
\usepackage{algorithmic}

\floatname{algorithm}{Procedure}

\usepackage{enumerate}

\usepackage{enumitem}

\DeclareMathOperator*{\argmax}{argmax}

\begin{document}
%
\title{A Hybrid Cache Architecture for Meeting Per-Tenant Performance Goals in a Private Cloud
}

\author{\IEEEauthorblockN{Taejoon Kim}
\IEEEauthorblockA{\textit{Computer Science Department} \\
\textit{Texas A\&M University}\\
Commerce, TX 75428\\
tkim3@leomail.tamuc.edu}
\and
\IEEEauthorblockN{Yu Gu}
\IEEEauthorblockA{ \textit{Department of Computer Science \& Engineering} \\
\textit{University of Minnesota}\\
Minneapolis, MN 55455 \\
 yugu@cs.umn.edu}
\and
\IEEEauthorblockN{Jinoh Kim}
\IEEEauthorblockA{\textit{Computer Science Department} \\
\textit{Texas A\&M University}\\
Commerce, TX 75428\\
jinoh.kim@tamuc.edu}
}

\maketitle

\begin{abstract} 

The in-memory cache system is an important component in a cloud for the data access performance.
As the tenants may have different performance goals for data access depending on the nature of their tasks, effectively managing the memory cache is a crucial concern in such a shared computing environment.
Two extreme methods for managing the memory cache are  {\em unlimited sharing} and  {\em complete isolation}, both of which would be inefficient with the expensive storage complexity to meet the per-tenant performance requirement.
%
In this paper, we present a new cache model that incorporates global caching (based on  unlimited sharing) and  static caching (offering  complete isolation) for a private cloud, in which it is critical to offer the guaranteed performance while minimizing the operating cost. 
This paper also presents a cache insertion algorithm tailored to the  proposed cache model. 
From an extensive set of experiments conducted on the simulation and emulation settings, the results confirm the validity of the presented  cache architecture and insertion algorithm showing the optimized use of the cache space for meeting the per-tenant performance requirement.
\end{abstract}


%
\IEEEpeerreviewmaketitle

\input{intro.tex}

\input{bg.tex}

\input{shared.tex}

\input{design.tex}

\input{eval.tex}

\input{conc.tex}


\let\OLDthebibliography\thebibliography
\renewcommand\thebibliography[1]{
  \OLDthebibliography{#1}
  \setlength{\parskip}{0pt}
  \setlength{\itemsep}{2pt plus 0.3ex}
}

\bibliographystyle{unsrt}
\bibliography{paper}

\end{document}

%% file: intro.tex
\section{Introduction}
	\label{sec:intro}

Caches are an important component of  computing systems for the data access performance.
A stand-alone system consists of a hierarchy of cache memories to reduce the gap between the speed of processors and memory.
The memory cache system utilizing a collection of memory space from the distributed computing resources 
has been widely deployed for large-scale data access for many data-intensive applications in diverse sectors including  eCommerce, financial, streaming, and scientific computing.
For example, the leading service providers such as Google, Facebook, and Amazon rely on memory caching using Memcached~\cite{Memcached} and Redis~\cite{Redis}, to accommodate a huge volume of data access transactions in a real-time manner.

In a cloud, the in-memory cache  system is an important component to preserve the quality of service for individual tenants with  different data access characteristics.
To estimate the data access performance, {\em cache hit rate} has been widely employed 
since the performance gap between cache memory and secondary storage (e.g., hard drives) is not comparable: 
the access latency to memory is 15--90ns, while it takes 100 us--15ms to access disks~\cite{xu2018real}.
This considerable discrepancy can result in a significant reduction of the access latency even with a small increase in cache hit rate;
for example, over 35\% of latency reduction can be expected only with 1\% increase in hit rate~\cite{Cliffhanger}.  
In this study, we use cache hit rate to specify the data access performance and requirement.

Optimizing  system resources is another important concern in a cloud.
That is, allocating the minimal cache space meeting the performance requirement should be desirable. 
Adding extra cache resources due to a non-optimized use of the memory cache is costly and inconvenient due to the possibility of service disruption for the installation and reconfiguration.
This is particularly important in a {\em private cloud}, in which an organization operates the cloud infrastructure in an isolated manner;
one of the important motivations of the private cloud is to maximize the utilization of existing on-premise resources~\cite{dillon2010cloud}. 
In enterprise private clouds, the total available resources are relatively limited compared to mega-scale public cloud. In addition, there are typically different tiers of applications with different data access performance requirements. 
This study targets on a private cloud where it is critical to offer the guaranteed performance, while minimizing the operating cost.

\begin{figure}[!tb]
 \centering
 \subfigure[Global caching] {
    \label{fig:global_cache}
    \includegraphics[width=.45\columnwidth]{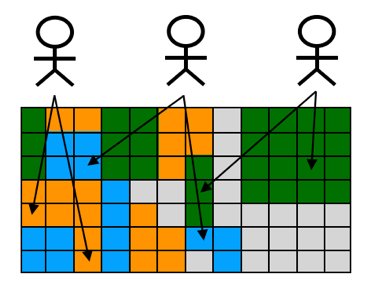}}
 \hspace{.1in}
 \subfigure[Static caching] {
    \label{fig:static_cache}
    \includegraphics[width=.45\columnwidth]{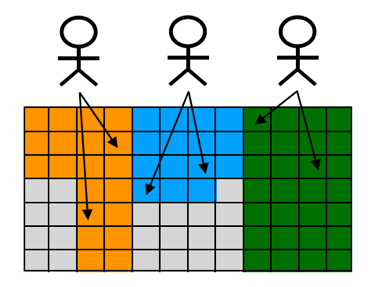}}
 \label{fig:cache_model}
 \caption{
Cache models:
global caching allows tenants to share the cache resources as a whole, whereas 
static caching allocates an isolated space to each user for the exclusive use.
}
\end{figure}

Two extreme methods for managing the memory cache are {\em unlimited sharing} and {\em complete isolation}, both of which would be inefficient with the expensive storage complexity to meet the per-tenant performance requirement.
The basic {\em global caching} scheme allows unlimited sharing of the cache space, 
and hence, it would be hard to guarantee the specified performance goal for data access for individual users (Fig.~\ref{fig:global_cache}).
In contrast, {\em static caching} offers an isolated cache space to each tenant for the exclusive use of the allocated cache space (Fig.~\ref{fig:static_cache}).
While straightforward to manage without sharing,
a critical downside of static caching is that it is not an easy task to estimate the required amount of the cache space in advance.
In addition, the time-varying property of the workload patterns often requires over-booking of the cache space to keep meeting the performance goal, which may cause the waste of cache resources due to  under-utilization.

{\em Dynamic caching} manages the cache space adaptively by responding to the data access pattern changes~\cite{Chockler:2010,ChocklerLV11,Dynacache}.
In other words, the size of cache space for a tenant is adjusted (i.e., grown or shrunk) over time by monitoring the performance measure (e.g., hit rate). 
It is thus essential to accurately predict data access patterns to determine the required size of the cache space for individual users.
Inadequate predictions will lead to the waste of the resources or the failure of the performance guarantee. 
There have been a body of studies to address this problem,
such as based on the approximation to concave functions~\cite{ChocklerLV11} and a learning-based prediction~\cite{choi2019},
but it is still a challenging problem to make the accurate estimation of access patterns with an acceptable error bound.

In this work, we take an approach that incorporates global caching (based on unlimited sharing) and static caching (offering complete isolation) 
for the optimized use of the cache space while satisfying the performance requirement for individual tenants.
The underlying idea  is that static caching is basically utilized to meet the minimal requirement, while the global caching scheme is employed for the use of cache resources efficiently.
We present a new cache model based on this idea with a cache insertion algorithm tailored to the proposed cache architecture. 
To validate, the extensive experiments have been conducted on the simulation and emulation settings, and the experimental results confirm the effectiveness of the presented cache architecture not only to meet the per-tenant requirement but also to optimize the cache space use.


The key contributions of this paper are as follows:

\begin{itemize}
   \item We set up a problem of multi-tenant caching with a two-level performance requirement to specify the quality of data access.
We discuss the limitations of global and static caching techniques with the experimental results obtained from the preliminary experiments. 
   \item 

We introduce two strategies for the cache insertion to efficiently manage the shared cache space in a tenant-aware manner:
(1) {\em fair-sharing}, which shares the cache space to make the cache hit rates of the active users to be equal.
(2) {\em selfish-sharing}, 
by which a user yields the occupied cache space {\em only if} she predicts the sharing does not make her hit rate below the performance requirement. 
   \item We present a new cache architecture incorporating the global and static caching schemes, which consists of two sections: {\em dedicated cache} (DC) and {\em shared cache} (SC). The former is the cache space partitioned for individual users and the latter is the shared space that the tenants compete each other to utilize.  
A cache insertion algorithm tailored to the proposed cache architecture will also be introduced.
   \item We implement a simulation model for extensive experiments with diverse parameter settings.
In addition, we build up an emulation environment on top of Apache Ignite~\cite{Ignite}, an in-memory caching framework, for evaluating in a real setting.
We report our experimental results conducted on these environments. 

\end{itemize}

The organization of this paper is as follows. 
In the next section, we present the problem description and the methods for experiments with the a summary of the closely related studies.
Section~\ref{sec:shared} presents a method to share the cache space among users with a simple, tenant-aware cache insertion algorithm.
In Section~\ref{sec:design}, we propose a new cache model incorporating the global and static caching to manage the cache in a way to fulfill the two-level performance requirement defined in the problem description, and report the experimental results conducted in the simulation settings.
We also report  the experimental results conducted on a real memory cache setting with Apache Ignite 
in Section~\ref{sec:eval}.
Finally, we conclude our presentation in Section~\ref{sec:conc} with a summary of the work and  future directions.

%% file: bg.tex
\section{Background}
	\label{sec:bg}

\subsection{Problem Description}
	\label{sec:prob}
	
\begin{table}[!tb]
\caption{Notations}
\label{tab:notation}
\centering
\small
\begin{tabular}{l|l}
\hline
Notation & Description \\
\hline
$C$ & Total cache capacity in the system \\
$U_k$ & User $k$ in the tenant set $U$ ($u_k \in U$) \\
$H_k$ & Minimal cache hit rate  for $u_k$ \\
	& (hard requirement)\\
$S_k$ & Desired cache hit rate  for $u_k$ \\
	& (soft requirement)\\
$h_k$ & Measured cache hit rate for $u_k$ \\
$g_k$ & $g_k=h_k - S_k$ \\
$G$   & $G=min(h_k - S_k)$ for all $k$ \\
\hline
\end{tabular}
\end{table}

\begin{figure}[!tb]
 \centering
 \subfigure[Global caching] {
    \label{fig:global_cache_ex}
    \includegraphics[width=.75\columnwidth]{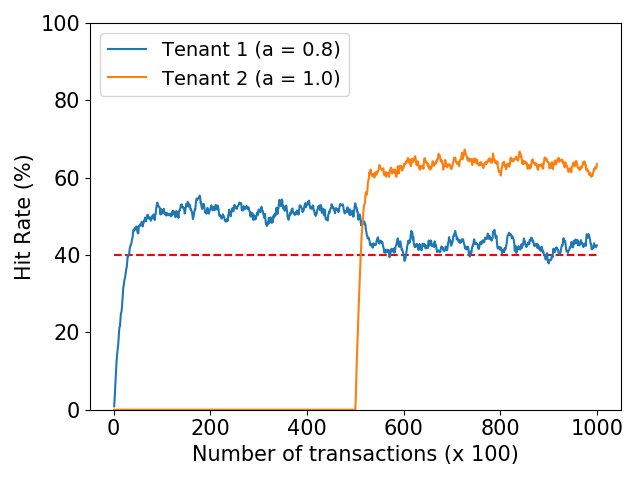}}
 \subfigure[Static caching] {
    \label{fig:static_cache_ex}
    \includegraphics[width=.75\columnwidth]{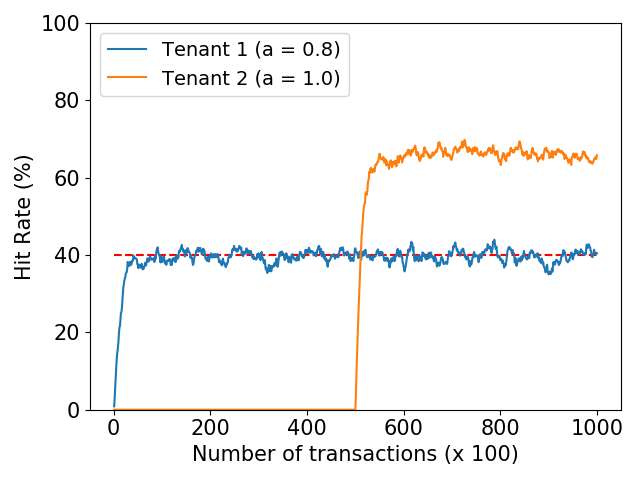}}
 \caption{Example of caching performance:
In this scenario, $u_1$ ($\alpha=0.8$) is alone in the system initially, and $u_2$ ($\alpha=1.0$) becomes active at around $t=50,000$ (number of transactions).
The figure shows that global caching barely meets the required hit rate (40\%), while static caching does not meet the requirement for $u_2$ due to the independent use of the isolated cache space among tenants.
 }
 \label{fig:cache_ex}
\end{figure}

We assume a multi-tenant cloud with a large-scale in-memory cache system (such as using Apache Ignite~\cite{Ignite} and Amazon ElastiCache~\cite{ElastiCache}).
Individual tenants have their own performance requirements for data access (chosen based on budgets, application-level requirements, etc).
To specify  the performance requirement, two levels (namely {\em hard}  and {\em soft} requirements) are often used for flexibility.
In real time systems, for example, the hard deadline must be guaranteed whereas the soft deadline would preferably be met~\cite{brandenburg2007}.   
The work in~\cite{moon2011} employed hard and soft SLAs for database scheduling requirements. 
Similarly, we define two levels to specify the performance requirement for individual users, as follows:

\begin{itemize}
	\item {\em Hard requirement}: the minimal cache hit rate that should be satisfied at any time for the tenant
	\item {\em Soft requirement}: the desired cache hit rate as long as the cache resources are currently sufficient (not mandatorily but preferably) 
\end{itemize}

The hard requirement is the lowest bound of the hit rate, and the measured hit rate for the tenant must be at least equal to this requirement,
whereas the soft requirement is the desired performance goal that would be satisfied unless the system suffers from the shortage of cache space for any reason (e.g., too many active concurrent users, a part of cache server failure, etc).
%
Using the notations in Table~\ref{tab:notation}, we define the following two objectives to meet the hard and soft  requirements: \\


	{\em Objective 1}: For any $k$, $h_k \ge H_k$ 

	{\em Objective 2}: Maximize $G = \min_k (h_k - S_k)$ for all $k$

\vspace{.1in}
The first objective states the hard requirement for individual users.
The second objective is for the soft requirement, and we want to maximize 
the minimal gap ($G$) between the measured hit rate and the desired hit rate.
If the minimal gap is a zero or any positive number ($G \ge 0$), it indicates that the soft requirement for everyone is satisfied;
otherwise (i.e., $G<0$), it implies that there exists at least one user who has a measured hit rate smaller than the desired one.
To explain, let us suppose two tenants with $S_1 = S_2 = 70\%$ (soft requirements) and 
$h_1=80\%$ and $h_2=65\%$ (current hit rates).
In this scenario, we get $G=-5\%$ (because of $g_1 = 10\%$ and $g_2 = -5\%$) and our goal is to maximize $G$ if possible.
If another setting produces $h_1=72\%$ and $h_2=70\%$, this case is more preferable since $G=0$, 
implying the soft requirements for both users are met.

Do the existing caching techniques work for these objectives? Fig.~\ref{fig:cache_ex} gives an answer to this question. 
The scenario of this experiment is as follows.
We assume two tenants: $u_1$ (in blue) is alone in the system initially, and $u_2$ (in orange) becomes active at around $t=50,000$ (with respect to the number of transactions).
The access patterns are based on a Zipfian distribution for each user, with $\alpha=1.0$  for $u_1$ and $\alpha=0.8$ for $u_2$, where $\alpha$ is the Zipfian parameter.
Since $t=50,000$, the equal number of transactions are created alternately for the two users.
In this experiment, the cache size is set to 3,000 fixed-length slots (note that the use of fixed-length slots is described in Section~\ref{sec:sim}).
Assume the hard requirement is set to 40\% for both users.

The global caching scheme allows the users share the whole capacity in cache ($C=3,000$). 
As soon as tenant 2 comes in, the cache space is being shared as in Fig.~\ref{fig:global_cache_ex}, and we can see that $u_1$ sometimes violates the performance requirement.
For static caching (Fig.~\ref{fig:static_cache_ex}), it is assumed that a half of the total cache space ($\frac{C}{2}=1,500$) is assigned to each user.
Since $u_2$ has a relatively low skewness in distribution, it shows around 40\% for hit rate violating the performance requirement, while 
$u_1$ shows a much higher hit rate ($>$60\%) 
with the same size of  cache space.

\subsection{Methods for Experiments}
	\label{sec:ex_method}
	
\paragraph{Data access patterns}
In this work, we employ Zipfian distributions to model data access patterns. 
A  body of studies reported Zipf-like patterns for data access through the measurement studies~\cite{Breslau1999WebCA,Gummadi:2003,cherkasova2004analysis,sun2013workload,Calzarossa:2016}. 
For example, Zipf ($\alpha=0.9$) was assumed to model the load distribution for caching~\cite{ramaswamy2005cache}, while $\alpha=0.99$ was used for the key-value request workload in a cloud~\cite{shue2012performance}. 
The work in~\cite{lu2007virtual} assumed $\alpha=1.0$ for the Zipf-like file accesses in their evaluation. 
In addition, YCSB~\cite{YCSB} is a benchmark tool widely employed for modeling cloud workloads
based on the Zipfian distribution.  
In this work, we  assume Zipf-based workloads to model data access patterns with a range of coefficient values ($0.7 \leq \alpha \leq 1.0$), chosen based on the observations made by previous measurement studies.

\paragraph{Measuring cache hit rates}
The reliable measurement of cache hit rate is  important since 
this measure is used to estimate the data access performance in this study.
It is possible to measure hit rates in a window basis, calculating $\frac{\text{\# cache hits}}{\text{\# cache accesses}}$, but there can be a high degree of fluctuation.
For reliability, we estimate hit rates using an exponentially weighted moving average (EWMA) as in TCP to estimate RTT (round-trip-time).
In detail, suppose the averaged cache hit rate up to the previous time interval $(i-1)$ is $R_{i-1}$, and the newly observed cache hit rate for the current interval is $R_c$, then the new estimation for cache hit rate is: 
$R_i = \alpha \cdot R_c + (1-\alpha) \cdot R_{i-1}$. 
Here, $\alpha$ defines the weights and we set it to $\alpha=0.125$, which is a typical value for TCP RTT estimation.

\paragraph{Simulation and emulation} \label{sec:sim}
We develop a simulation model that implements the proposed architecture with the insertion methods, as well as the existing caching techniques including global caching and static caching as the baseline to compare.
To evaluate on a real setting, Apache Ignite~\cite{Ignite} is installed as the in-memory cache system with Yardstick (a benchmarking tool for Apache Ignite). 

\paragraph{Fixed cache item size} \label{sec:slab}
For simplicity, we assume that the memory cache system consists of a set of cache slots with the identical length.
Thus, we use ``number of slots'' to indicate the capacity of cache space (rather than a specific size such as using gigabytes). 
We leave the consideration of the slab design~\cite{Dynacache} to support multiple classes for cache slots as one of our future tasks.

\subsection{Related Work}
	\label{sec:related}
	
This section provides a summary of the previous studies related to our work.

There exist several past studies that investigated multi-tenant caching in a cloud based on the dynamic prediction approach. 
The work in \cite{Chockler:2010}
estimates cache hit rates based on a least-square fit
using a log function ($y = a + b\log x$).
Here, the output ($y$) is the estimated hit rate for the given cache size ($x$).
The authors also proposed a cache space utility model based on 
the stack distance 
to improve the overall cache hit rate in a cloud~\cite{ChocklerLV11}.
A recent study in \cite{choi2019}, the authors a prediction method using machine learning, to estimate the proper cache size for the workload in question.
The primary focus in dynamic caching is thus to minimize the prediction error, but it would be hard due to too many variables to estimate  workload patterns in advance.
Cliffhanger~\cite{Cliffhanger} employs an iterative method to respond to workload changes and incrementally optimizes the cache resource allocations in a reactive manner.
A latest work in~\cite{wu2019autoscaling} also provides a function of autoscailing to dynamically adjust the cloud storage in response to workload changes.

Dynacache~\cite{Dynacache} also estimates cache hit rates using stack distances.
This work proposes an approximation technique using  buckets to decrease  the computational complexity  for calculating stack distances.
Despite of the approximation, it would  still be expensive when assuming hundreds of concurrent applications.
This past work assumes multiple slab classes for cache allocation, each of which has a different size, to serve different types of application in a flexible manner.
In this study, we assume a fixed length for cache slots for simplicity and leave this question for our future exploration. 

FairRide~\cite{FairRide} and its successor~\cite{yu2018opus} attempt to address a problem of cache sharing among multiple users, but unlike our work, this work focuses ons a cheating problem by greedy users.
In this problem, a greedy user could deceive a system to obtain more cache space than others in an unfair manner.
Our work also assumes the sharing of cache space but the purpose of the sharing is to optimize the performance among users.

TinyLFU~\cite{TinyLFU} introduces an approximation method to implement the LFU (Least Frequently Used) policy, which is known to be the best with respect to cache hit rates.
To implement an approximated version of LFU, the authors employ a sketch data structure to count the frequency of data access, which is used to determine the items inserted to
the cache (i.e., cache admission control).
We are also interested in developing admission policies to efficiently share the cache space among tenants with the performance requirement.

The Segmented LRU (SLRU) inspired us to design a hybrid architecture for memory cache systems~\cite{SLRU}. 
This past work manages two distinctive segments in the computer cache, a probation segment and a protected segment, to capture recent popularity to implement the LFU policy.
A new cache item is inserted into the probation segment, which can be promoted in case of repeated accesses and the item moves to the protected area.  
If cache eviction is required in the protected segment, the evicted item is moved back to the probation area.
By doing so, SLRU allows temporarily popular items to stay longer in the cache than less popular ones.
The proposed hybrid cache architecture consists of two partitions in the cache similar with SLRU, but the purposes are different and our design is to meet the per-tenant performance requirement in a cloud (rather than implementing an LFU approximation).

%% file: shared.tex
\section{Tenant-aware Cache Sharing}
	\label{sec:shared}

In this section, we discuss how to share the given shared cache space in a tenant-aware manner, which
is connected to the second objective: Maximize $G = \min(h_k - S_k)$ for all $k$, described in the previous section.
Indeed, this objective is one of the {\em Max-Min} problems.
We address this problem by designing a cache insertion algorithm, as follows.

\begin{algorithm} 
 \caption{{\em Max-Min} cache insertion algorithm}
 \label{alg:evict1}
 \begin{algorithmic}
 \STATE {\bf Input}: a new cache item $c_i$ for $u_i$; \\
 \If{cache hit}{
   \Return;
 }
 \If{cache not full}{
   \STATE insert $c_i$ to an empty slot in the cache; \\
   \Return;
 }
 \STATE $X = \{\text{all active tenant IDs}\}$; \\
 \STATE $j = \argmax_{k \in X} (g_k = h_k-S_k)$ \\
 \STATE Evict one of the $u_j$'s cache items; \\
 \STATE Insert $c_i$ to the evicted slot; \\ 
 \end{algorithmic}
\end{algorithm}
 
\begin{figure}[!tb]
\centering
\includegraphics[width=0.75\columnwidth]{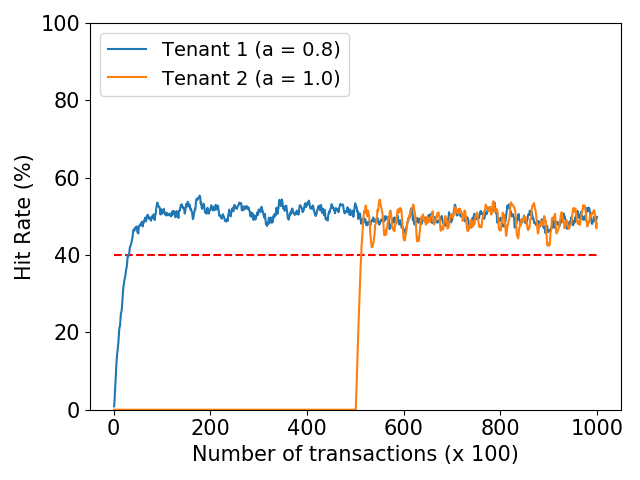}
\caption{
Example of caching based on Alg.~\ref{alg:evict1}:
The {\em Max-Min} sharing allows the users to share the cache space based on the measured hit rates, and the result shows both tenants meet the requirement. 
The experimental setting is identical to 
what is used in Fig.~\ref{fig:cache_ex}.
}
\label{fig:our_cache_ex}
\end{figure}

\begin{figure*}[!tb]
 \centering
 \subfigure[Cache hit rate (global caching)] {
    \label{fig:global_hr}
    \includegraphics[width=.38\textwidth]{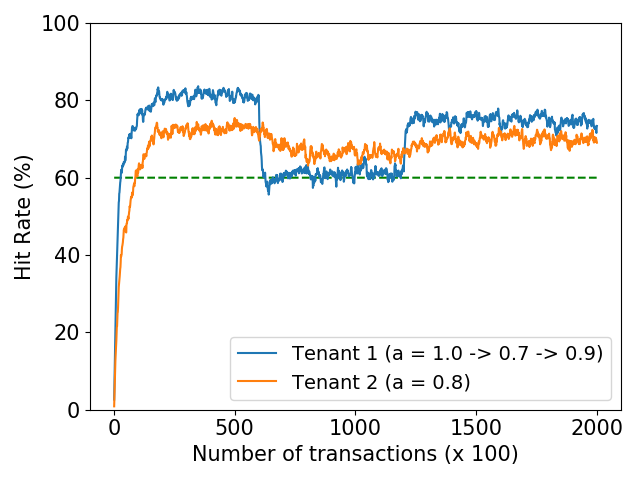}}
 \hspace{.5in}
 \subfigure[Cache space usage (global caching) ] {
    \label{fig:global_usage}
    \includegraphics[width=.38\textwidth]{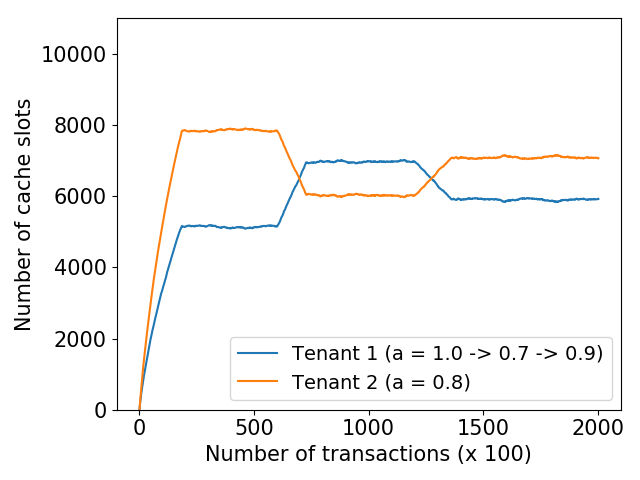}}
 \subfigure[Cache hit rate (fair-sharing)] {
    \label{fig:fair_hr}
    \includegraphics[width=.38\textwidth]{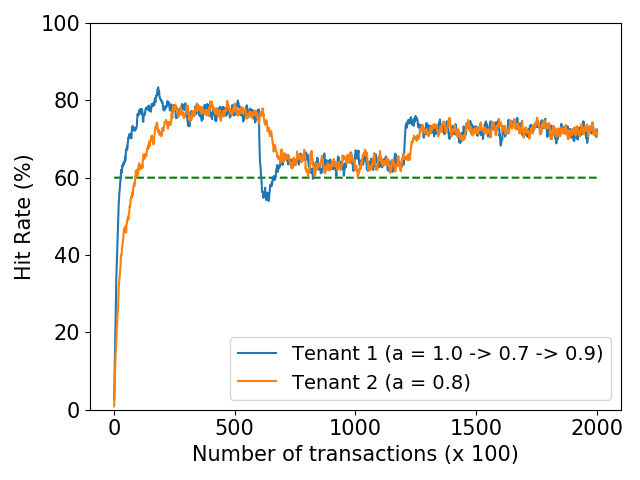}}
 \hspace{.5in}
 \subfigure[Cache space usage (fair-sharing)] {
    \label{fig:fair_usage}
    \includegraphics[width=.38\textwidth]{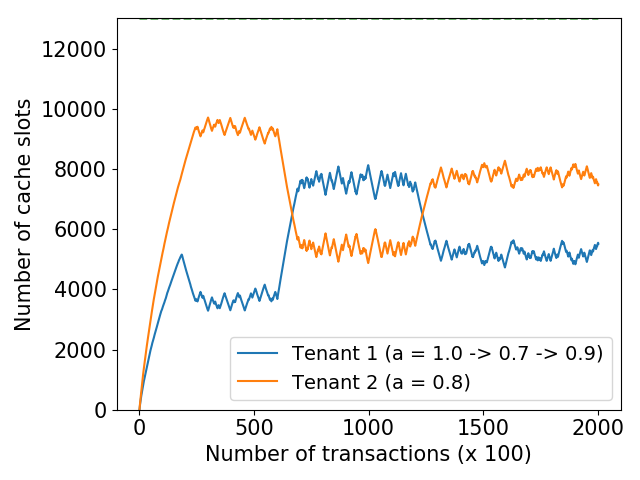}}
 \caption{
Comparison between global sharing and fair-sharing (\# cache slots = 13,000):
(a) and (b) show the hit rate and cache space usage when using global sharing, and (c) and (d) show the results for fair-sharing.
Fair-sharing allows a user with a lower skewness to occupy more cache space to manage the performance to be equal.  
 }
 \label{fig:shared_2tenants}
\end{figure*}

\begin{figure*}[!tb]
 \centering
 \subfigure[Cache hit rate (selfish-sharing)] {
    \label{fig:self_hr}
    \includegraphics[width=.38\textwidth]{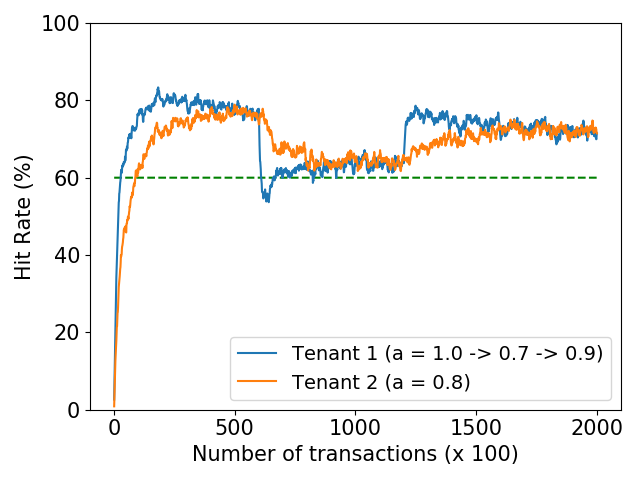}}
 \hspace{.5in}
 \subfigure[Cache space usage (selfish-sharing)] {
    \label{fig:self_usage}
    \includegraphics[width=.38\textwidth]{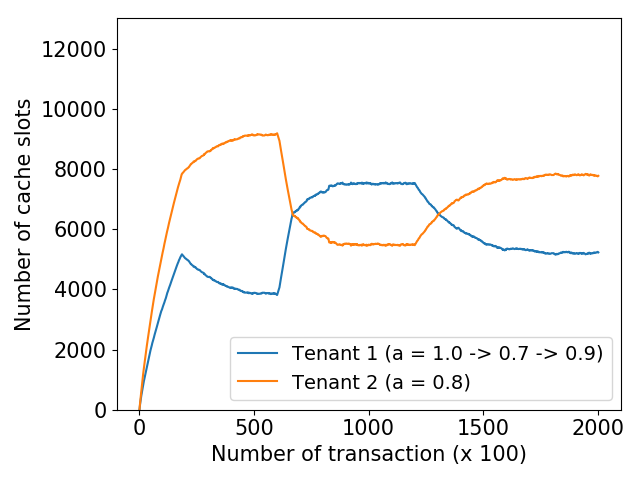}}
 \caption{
Cache performance and usage for selfish-sharing (\# cache slots = 13,000):
Selfish-sharing works differently than fair-sharing. Tenant 2 is not willing to give up the cache space it occupies to manage the performance above the requirement.
 }
 \label{fig:self_2tenants}
\end{figure*}

\begin{figure}[!tb]
 \centering
 \subfigure[Two tenants] {
    \label{fig:TwoTenantsSlots}
    \includegraphics[width=.38\textwidth]{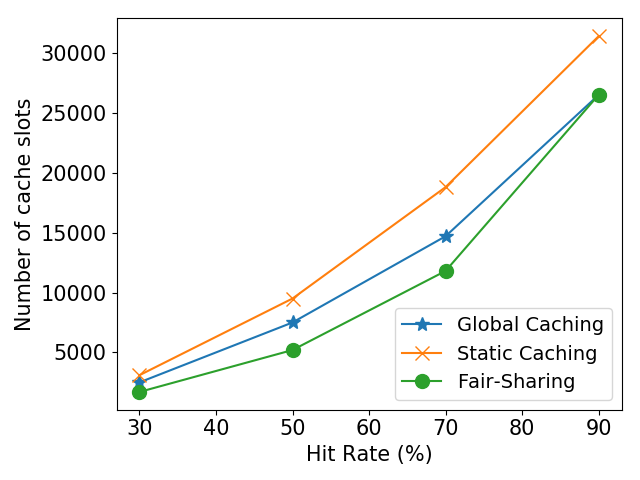}}
 \subfigure[Four tenants] {
    \label{fig:FourTenantsSlots}
    \includegraphics[width=.38\textwidth]{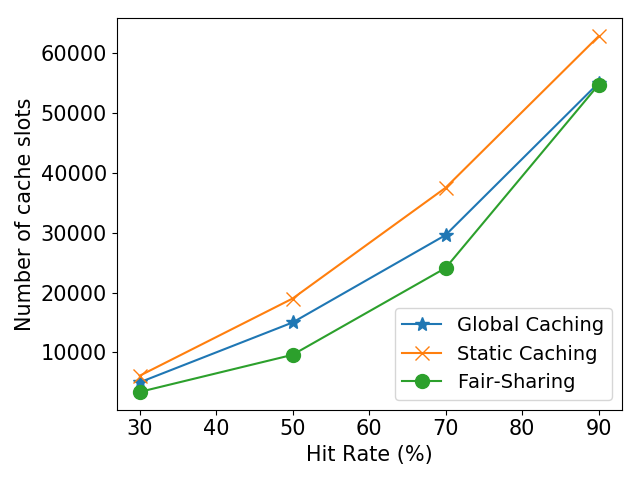}}
 \caption{
Comparison of space complexity:
(a) a two-user setting  with $\alpha=1.0$ and $\alpha=0.7$, respectively, and (b) a four-user setting with $\alpha=0.7$, $\alpha=0.8$, $\alpha=0.9$ and $\alpha=1.0$, respectively.
Fair-sharing optimizes the usage of cache space saving up 36\% compared to global caching and up to 49\% compared to static caching.
The benefits are greater when the required hit rate ($x$\%) is smaller, particularly when comparing to global caching. 
 }
 \label{fig:RequiredSlots}
\end{figure}


Alg.~\ref{alg:evict1} illustrates our implementation to solve the {\em Max-Min} problem. 
For a newly arrived item, if the item is already in the cache (``cache hit'') or if the cache has any empty spot, the algorithm works the same as the existing technique;
otherwise, our technique first searches the user who currently shows the greatest gap between the measured hit rate ($h_k$) and the desired hit rate ($S_k$).
A cache item occupied by the user with the greatest gap is selected to be evicted,
and then the cache slot is given to the new item.
Note that we do not assume any specific replacement strategy to determine what is evicted, but it can be any such as LRU and FCFS.
Fig.~\ref{fig:our_cache_ex} is an example of caching based on Alg.~\ref{alg:evict1} and  compares to Fig.~\ref{fig:cache_ex}.
From the figure, the two users show almost the identical hit rates by sharing the cache space and neither violates the defined  performance requirement unlike the global and static caching techniques.

To see how the algorithm works more in detail, we set up an experiment
with two users: tenant 1 with the varying access patterns ($u_1$) and tenant 2 with the static pattern ($u_2$).
Specifically, $u_2$ has an unchanged access pattern ($\alpha=0.8$), whereas $u_1$ shows $\alpha=1.0$ initially, $\alpha=0.7$ next at $t=600K$, and $\alpha=0.9$ at around $t=1200K$ finally.
We assumed the equal number of transactions made by the users,
and this assumption will be applied for the rest of the experiments unless otherwise mentioned.
The performance requirement for the two tenants has been set to 60\% (of hit rate).

Fig.~\ref{fig:shared_2tenants} 
shows the experimental result with respect to cache hit rate and space usage for global sharing and our sharing method described in Alg.~\ref{alg:evict1} (called ``fair-sharing''). 
As discussed, global sharing performs in a tenant-unaware fashion, and 
Fig.~\ref{fig:global_hr} shows the hit rate changes over time:
$u_1$ shows a large degree of fluctuations when it experiences the access pattern changes.
Compared to this, $u_2$ keeps the hit rate with a minor variation.
Fig.~\ref{fig:global_usage} plots the cache space usage over time (with respect to the number of cache slots).
%
Fig.~\ref{fig:fair_hr} and Fig.~\ref{fig:fair_usage} show how Alg.~\ref{alg:evict1} performs under the same condition.
From Fig.~\ref{fig:fair_hr}, the two users produce almost the identical hit rates over time by managing the cache space as shown in
Fig.~\ref{fig:fair_usage}, in which $u_1$ utilizes more space in the middle of the time than global sharing in Fig.~\ref{fig:global_usage}.
We can see that fair-sharing meets the performance requirement by active sharing.

One may not prefer this type of sharing (fair-sharing); rather she may want more a selfish way in sharing, with which a user yields the occupied cache space {\em only if} the current performance is higher than the defined requirement.
We call this strategy ``selfish-sharing.''
%
An essential element to implement selfish-sharing is the prediction of performance 
 in case of the loss of a certain number of cache slots occupied by the user.
If the predicted performance is still acceptable, the number of slots can be taken from the user and given to another.
The prediction takes place at every $N$ transactions with the simple assumption of the maximum number of loss is $N$. 
Our implementation for selfish-sharing is based on a linear regression function.


Fig.~\ref{fig:self_2tenants} shows how selfish-sharing performs under the same setting as in Fig.~\ref{fig:shared_2tenants}.
In Fig.~\ref{fig:self_hr}, both users meet the performance requirement,
although the pattern is slightly different.
As can be seen from Fig.~\ref{fig:self_usage}, the resulted lines are smooth  although the overall pattern is almost the same as fair-sharing.
This is because selfish-sharing determines the sharing of the occupied slots based on prediction. 
In our experiment, the prediction takes place at every $N=100$ transactions, by which a tenant estimates the hit rate for the next cycle with the assumption of the loss of 100 slots.
If the estimated hit rate still meets the requirement for a tenant, it operates in a same way as fair-sharing; otherwise, selfish-sharing rejects the sharing request from other tenants who want to utilize the slots occupied by that tenant.

We next examine the space complexity, by measuring how much cache space is required to meet $x$\% hit rate for the active tenants.
We design two independent experiments: (1) a two-user setting  with $\alpha=1.0$ and $\alpha=0.7$, respectively; and (2) a four-user setting with $\alpha=0.7$, $\alpha=0.8$, $\alpha=0.9$ and $\alpha=1.0$, respectively.
Fig.~\ref{fig:RequiredSlots} shows the required cache space to meet $x = 30\%-90\%$ for the two settings.
Note that we assume the equal space allocation to individual users for static caching.
As can be seen from the figure, fair-sharing optimizes the usage of cache space saving up 32\% compared to global caching and up to 45\% compared to static caching for the two-tenant setting (Fig.~\ref{fig:TwoTenantsSlots}). 
With four tenants, the result is somewhat similar with the two-tenant setting and fair-sharing shows up to 36\% and 49\% compared to global and static caching, respectively.
The experimental result shows greater benefits when the required hit rate ($x$\%) is smaller, particularly when comparing to global caching. 
We omit the result of selfish-sharing since its space requirement should be the same as fair-sharing for the specified hit rate.

%% file: design.tex
\section{Proposed Cache Architecture}
	\label{sec:design}

\begin{figure}[!tb]
\centering
\includegraphics[width=1.0\columnwidth]{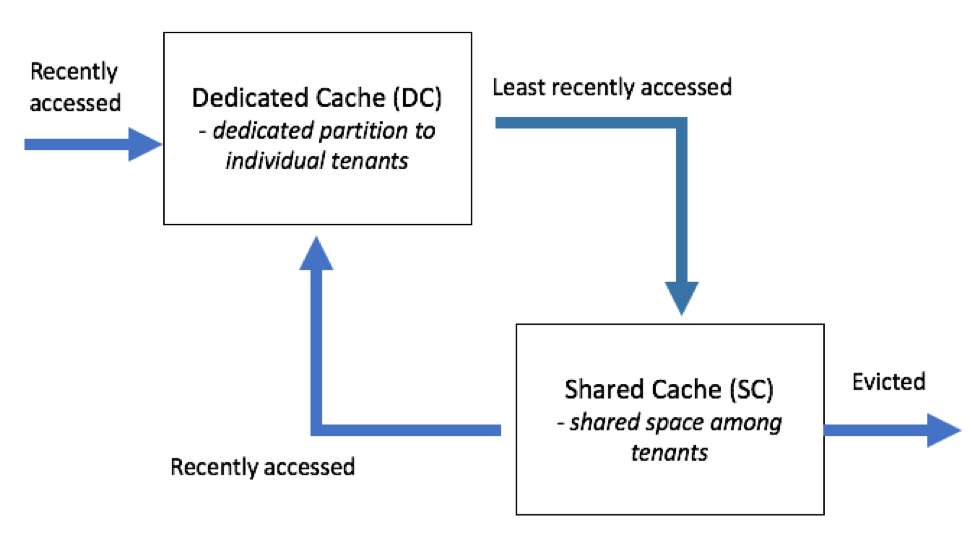}
\caption{
Proposed cache architecture:
The new cache architecture consists of two components of dedicated cache ($DC$) and shared cache ($SC$).
$DC_k$ is an isolated space for tenant $k$ to meet the hard requirement, while 
$SC$ is a shared space among tenants and the users compete each other to utilize this partition.
}
\label{fig:hybrid_arch}
\end{figure}

In this section we present a new cache architecture that incorporates static and global caching,
to attain the objectives discussed in Section~\ref{sec:prob}.
Fig.~\ref{fig:hybrid_arch} illustrates the proposed cache architecture with two components in the memory cache system, as follows:

\begin{itemize} 
  \item {\em Dedicated Cache (DC)}, which is composed of $N$ partitions and a partition ($DC_k$) is allocated to a single tenant ($u_k$). This space is not shared among tenants and used in a mutually exclusive way.
  \item {\em Shared Cache (SC)}, which is shared among tenants, and the tenants compete each other to utilize this shared space. 
\end{itemize} 

A $DC$ partition is utilized by a single user only, and hence, this space is assigned to meet the ``minimal'' performance requirement ({\em Objective 1}), while $SC$ is a shared space to achieve the ``desired'' performance goal ({\em objective 2}).
The user $u_i$ utilizes $DC_i$ first and then $SC$ in case of no more empty slot available in the dedicated  partition. 
As shown in Fig.~\ref{fig:hybrid_arch}, hot items (accessed recently) are stored in $DC$, while cold items (accessed a long time ago) are kept in $SC$ since another user ($u_j$) possibly takes up the slot in $SC$ currently occupied by $u_i$ at any time as a result of the competition.
%

\begin{algorithm} 
\small
 \caption{{\em Cache insertion algorithm for the proposed cache architecture ({\em fair-sharing})}
 }
 \label{alg:evict2}
 \begin{algorithmic}
 \STATE {\bf Input}: a new cache item $c_i$ for $u_i$; \\
 \If{Hit from $DC_i$}{
   \Return;
 }
 \If{$DC_i$ is not full}{
   \STATE Insert $c_i$ to an empty slot in $DC_i$; \\
   \Return;
 }
 \If{Hit from $SC$}{
   \STATE Find a victim from $DC_i$; \\
   \STATE Swap the hit slot in $SC$ and the victim slot in $DC_i$; \\
   \Return;
 }
 \If{$SC$ is not full}{
   \STATE Insert $c_i$ to $SC$; \\
   \STATE Find a victim from $DC_i$; \\
   \STATE Swap the inserted slot in $SC$ and the victim slot in $DC_i$; \\
   \Return;
 }
 \STATE $X = \{\text{all active tenant IDs}\}$; \\
 \STATE $j = \argmax_{k \in X} (g_k = h_k-S_k)$; \\
 \While{true}{
   \eIf{$u_j$ occupies any slot in $SC$}{
     \STATE Find a victim from $SC$ occupied by $u_j$; \\
     \STATE Evict the victim;
     \STATE Insert $c_i$ to the victim slot in $SC$; \\
     \STATE Find a victim from $DC_j$; \\
     \STATE Swap the inserted slot in $SC$ and the victim slot in $DC_j$; \\
     \STATE break; \\
   }{
     \STATE $X = X - \{j\}$; \\
     \STATE $j = \argmax_{k \in X} (g_k = h_k-S_k)$; \\
   }   
 }
 \end{algorithmic}
\end{algorithm}

Alg.~\ref{alg:evict2} illustrates an algorithm for cache insertion tailored to the proposed cache architecture.
As described, $DC$ keeps hot items recently accessed while $SC$ stores relatively old items to allow the eviction from $SC$ in case of cache full. 
For a new cache item $c_i$ for $u_i$, if the new item is found from $DC_i$ or $DC_i$ has an empty slot, the algorithm simply works with $DC_i$ and the insertion procedure is over;
otherwise, $SC$ is next considered to complete the insertion.
If $SC$ has an empty slot, $c_i$ is inserted to the empty slot; if $SC$ is already full, a victim should be identified to make a room in $SC$ for the new item.
The victim in $SC$ is chosen based on the gap $G$ as in Alg.~\ref{alg:evict1}, and a slot in SC owned by the victim is assigned to the new item to be inserted.    
After the insertion in $SC$, a swapping takes place between $DC_i$ and $SC$ to keep the hot item in the dedicated cache: 
a victim item is chosen from $DC_i$ that is moved to the slot in $SC$ occupied by the new item, and the new item is stored in the victim slot in $DC_i$.

Alg.~\ref{alg:evict2} is for the fair-sharing strategy.
The only difference from selfish-sharing is the $while$-loop condition, located in the middle of the algorithm.
In the $while$-loop, fair-sharing finds a victim from the tenant currently having the greatest gap $G$.
In contrast, selfish-sharing first checks whether $G > 0$ or not, and the tenant is {\em not} selected as a victim in case of $G \le 0$.

\begin{figure*}[!tb]
 \centering
 \subfigure[Global caching] {
    \label{fig:global_5000_5to1}
    \includegraphics[width=.32\textwidth]{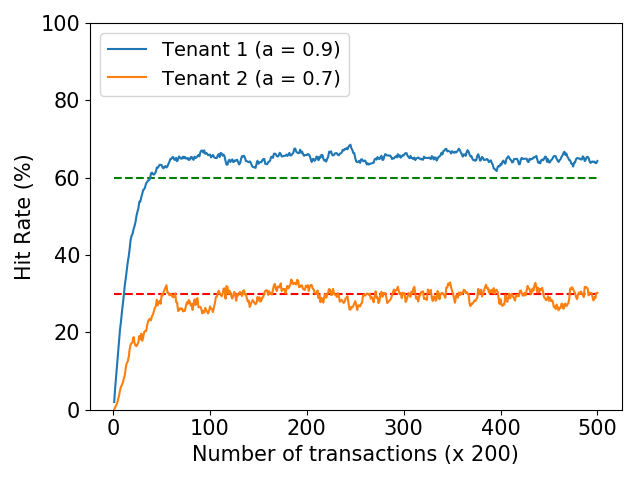}}
 \subfigure[Static caching] {
    \label{fig:static_5000_5to1}
    \includegraphics[width=.32\textwidth]{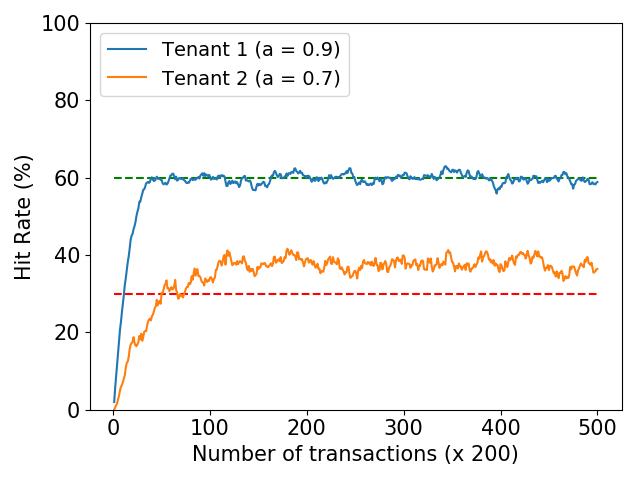}}
 \subfigure[Fair-sharing (Setting 1)] {
    \label{fig:fair_1000_1000_3000_5to1}
    \includegraphics[width=.32\textwidth]{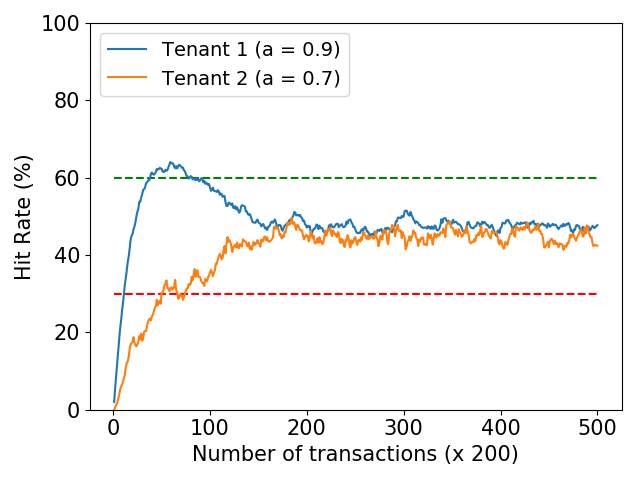}}
 \subfigure[Selfish-sharing (Setting 1)] {
    \label{fig:selfish_1000_1000_3000_5to1}
    \includegraphics[width=.32\textwidth]{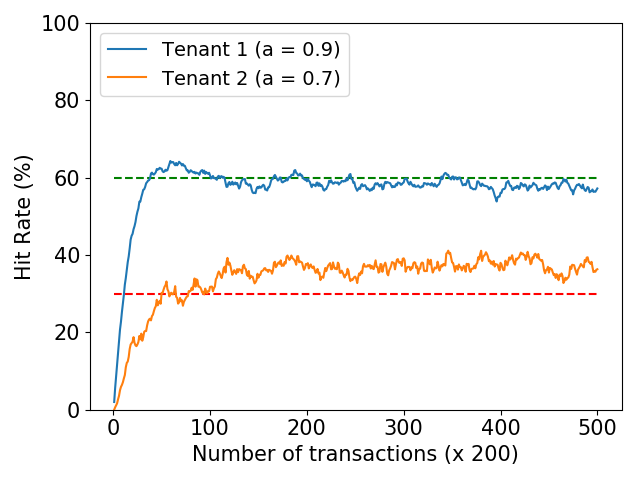}}
 \subfigure[Fair-sharing (Setting 2)] {
    \label{fig:fair_2000_2000_1000_5to1}
    \includegraphics[width=.32\textwidth]{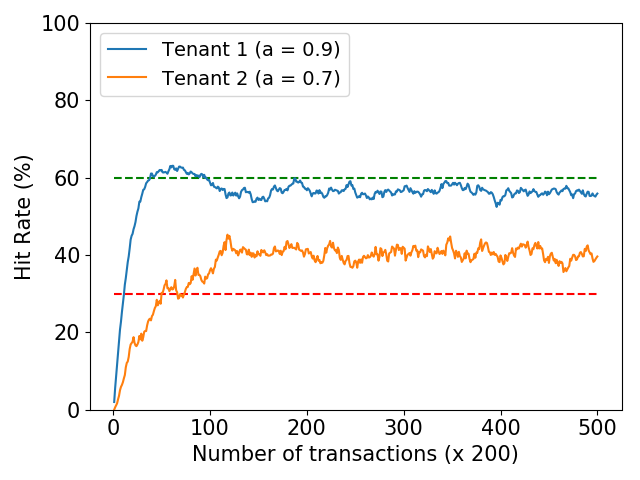}}
 \subfigure[Selfish-sharing (Setting 2)] {
    \label{fig:selfish_2000_2000_1000_5to1}
    \includegraphics[width=.32\textwidth]{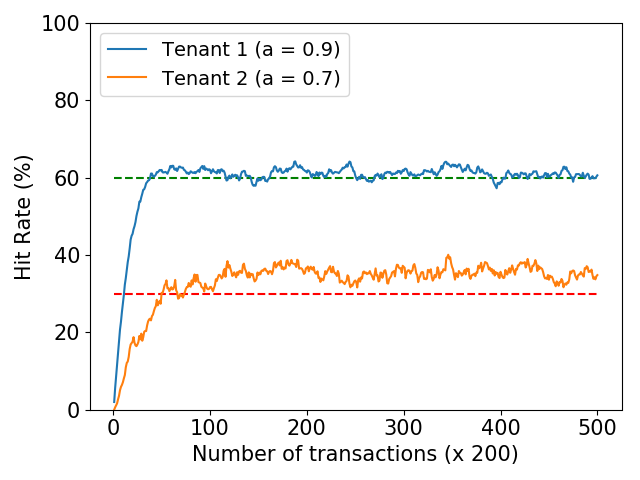}}
 \caption{
Experimental result on the proposed cache architecture (\# cache slots = 5,000):
(a) with global caching, $u_1$ utilizes more space with a greater number of transactions, which makes $u_2$ to be in trouble, not meeting even the hard requirement;
(b) static caching exploits the isolation of the cache space among tenants,
but this isolation might result in a worse scenario that a tenant shows a very high hit rate while the others suffer from the performance violation;
(c) \& (e)  
fair-sharing produces quite different results for the different settings, and 
the greater $SC$ size in Setting 1 shows a balance between the users with respect to hit rate;  
(d) \& (f) selfish-sharing does not show much differences for the two settings because the cache resources are not sufficient to meet the $u_1$'s soft requirement and $u_1$ does not offer her slots in $SC$ to $u_2$ since $u_2$'s hit rate is still above the hard requirement.
}
 \label{fig:cache_5000_5to1}
\end{figure*}

We now examine  how the new insertion algorithm works.
To see that, we assume the following setting:

\begin{itemize}
  \item Total number of slots in the cache = 5,000
  \item Hard/soft  requirement: $H=30\%$ and $S=60\%$
  \item Workload distribution: $u_i$ ($\alpha$=0.9) and $u_2$ ($\alpha$=0.7)
  \item Injection rate: $u_1$:$u_2$ = 5:1 
  \item Cache organization: 
	\begin{itemize}
		\item $|DC_k|$=1,000 and $|SC|$=3,000 (``Setting 1'')
		\item $|DC_k|$=2,000 and $|SC|$=1,000 (``Setting 2'')
	\end{itemize}
\end{itemize}

We assume a scenario with an insufficient cache size, 
and it is {\em not} possible to meet the soft requirement in this setting. 
The injection rate is the ratio between the users with respect to the number of transactions, and we assume an asymmetric rate of 
5:1, in which $u_1$ creates 5X greater number of transactions than $u_2$.
We experimented with two different cache organizations for  DC and SC as described above.
The selection of $DC_k$ size is one of the important tasks required for the further exploration in the future, and a simple option is to refer to the required number of slots to meet the hard requirement ($H$) under the assumption of the least skewed distribution (e.g., $\alpha=0.7$ in our setting) for the entire users.
We compare the experimental result with the existing caching methods, and assume each tenant has the equal partition size for static caching. 

Fig.~\ref{fig:cache_5000_5to1} shows the result of the experiment. 
With global caching, $u_1$ utilizes more space with a greater number of transactions, which makes $u_2$ to be in trouble, not meeting even the hard requirement, which is a critical violation for quality of service.
Static caching exploits the isolation of the cache space among tenants;
however, the isolation might result in a worse scenario that a tenant yields a very high hit rate ({\em over-provisioning}) while the others suffer from the performance violation ({\em under-provisioning}), as discussed in Fig.~\ref{fig:static_cache_ex}.

The plots from Fig.~\ref{fig:fair_1000_1000_3000_5to1} to Fig.~\ref{fig:selfish_2000_2000_1000_5to1} show the results with our management techniques with two different settings for $DC$ and $SC$ sizes.
Fair-sharing produces quite different results for the different settings, and 
the greater $SC$ size in Setting 1 shows a balance between the users with respect to hit rate.  
In contrast, selfish-sharing does not show much differences for the two settings because the cache resources are not sufficient to meet the $u_1$'s soft requirement and $u_1$ does not offer her slots in $SC$ to $u_2$ since $u_2$'s hit rate is still above the hard requirement.

Note that we also conducted an experiment with the 1:1 injection rate, and observed that the caching strategies work similarly with a minor difference from the 5:1 injection rate.

%% file: eval.tex
\begin{figure*}[!tb]
 \centering
 \subfigure[Global caching] {
    \label{fig:global_11000_ignite_soft}
    \includegraphics[width=.32\textwidth]{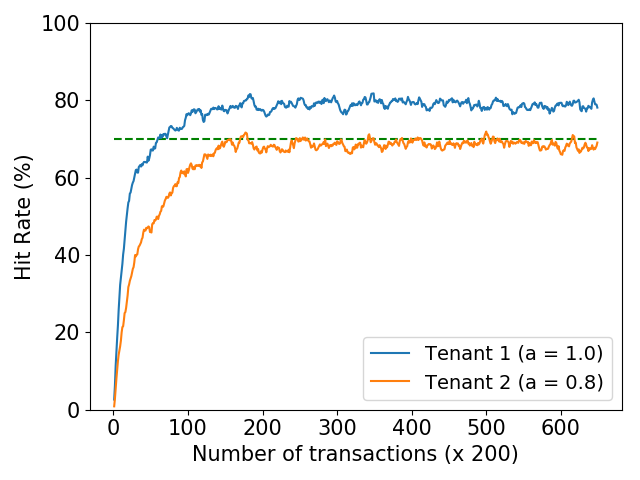}}
 \subfigure[Static caching] {
    \label{fig:static_11000_ignite_soft}
    \includegraphics[width=.32\textwidth]{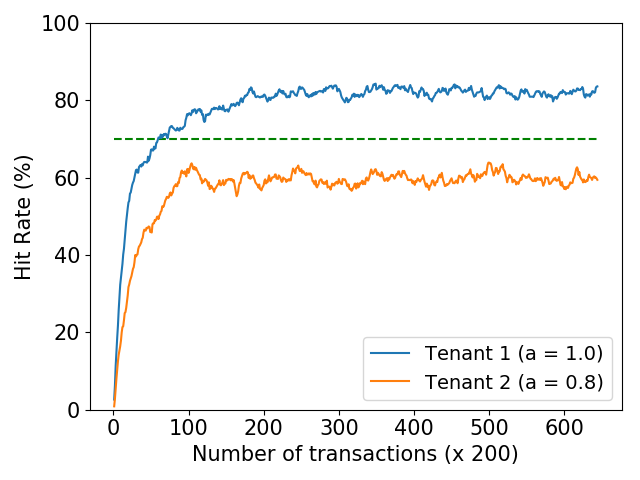}}
 \subfigure[Fair-sharing] {
    \label{fig:fair_11000_ignite_soft}
    \includegraphics[width=.32\textwidth]{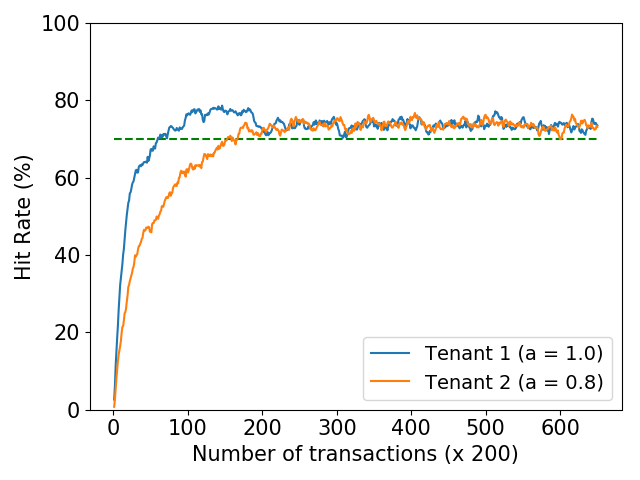}}
 \caption{
Experimental result on Apache Ignite with a single performance requirement (\# cache slots = 11,000):
(a) and (b) show that the existing techniques may not guarantee the performance requirement, while 
(c) fair-sharing meets the performance requirement by efficiently sharing.
 }
 \label{fig:ignite_11000_soft}
\end{figure*}

\begin{figure*}[!tb]
 \centering
 \subfigure[Fair-sharing] {
    \label{fig:fair_13000_ignite_hard_3tenants}
    \includegraphics[width=.38\textwidth]{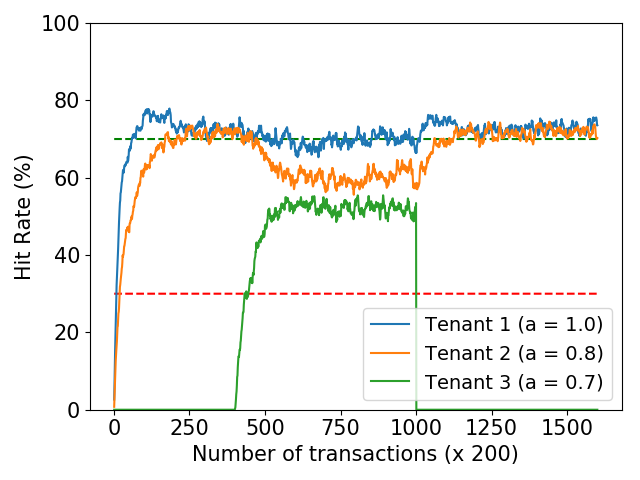}}
 \hspace{.5in}
 \subfigure[Selfish-sharing] {
    \label{fig:selfish_13000_ignite_hard_3tenants}
    \includegraphics[width=.38\textwidth]{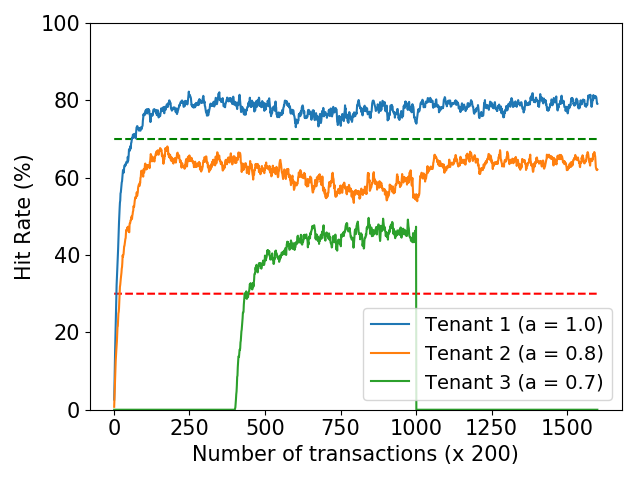}}
 \caption{
Experimental result on Apache Ignite with a two-level performance requirement (\# cache slots = 13,000):
while the two techniques meet the hard requirement,
$u_1$ and $u_2$ yields the occupied cache slots to $u_3$ in fair-sharing but 
$u_1$ does not give the occupied slots to the other two to manage its performance in selfish-sharing.
 }
 \label{fig:ignite_13000_hard_3tenants}
\end{figure*}

\section{Experiments on Apache Ignite}
	\label{sec:eval}

In this section, we report our experimental results conducted on a real in-memory cache system equipped with Apache Ignite~\cite{Ignite}.
We first describe the experimental setting in Section~\ref{sec:eval_setting}, and then discuss the experimental results performed with a single performance requirement in Section~\ref{sec:eval_soft}) and with the two-level requirements (hard and soft) in Section~\ref{sec:eval_both}.

\subsection{Experimental setting}
	\label{sec:eval_setting}

%
We conducted our experiments in an isolated setting with three computing systems connected over a fast Ethernet. The computing system consists of 4 CPU cores, 16 GB memory, and 1 TB hard disk drives. 
We installed Apache Ignite and MySQL as the in-memory cache infrastructure and the backend DBMS, respectively. 
Three nodes were configured for the in-memory cache service, each of which is configured with 4 GB cache space, and hence, the total cache space is 12 GB, in the system. Apache Ignite provides a set of built-in eviction methods including FCFS and LRU, and we simply chose LRU for cache replacement for DC and SC. 
One node is dedicated as a backend database server.
We ported the implementation of the cache insertion algorithm (presented in the previous section) onto  the yardstick benchmark tool.

\subsection{Experimental result with a single performance requirement}
	\label{sec:eval_soft}

We designed an experiment with a cache size that can meet the given performance requirement if the cache space is shared appropriately.
In this experiment, we allocate 11,000 slots to the cache system to meet 70\% hit rate as the single performance requirement.
The workload distributions are $\alpha=1.0$ and $\alpha=0.8$ for the active tenants, respectively.
Fig.~\ref{fig:ignite_11000_soft} shows the result and we can see that the existing techniques do not meet the performance requirement.
In contrast, fair-sharing guarantees the performance requirement by efficiently sharing the cache space.
Although not shown, we observed that selfish-sharing works as  intended satisfying the performance requirement for both tenants.

\subsection{Experimental result with a two-level performance requirement}
	\label{sec:eval_both}

We next assume three active tenants with the two-level performance requirements,
which are set to $H=30\%$ and $S=70\%$.
We assume different workload patterns for the tenants: $u_1$ with $\alpha=1.0$, $u_2$ with $\alpha=0.8$, and $u_3$ with $\alpha=0.7$.
In this scenario, we assume that two tenants are active over the entire time, while $u_3$ becomes active in the middle and leaves the system shortly.
The number of slots allocated is 2,000 slots for each $DC_k$ and 7,000 slots for $SC$. 

Fig.~\ref{fig:ignite_13000_hard_3tenants} demonstrates how the two proposed techniques work under the experimental setting described above.
The plots in Fig.~\ref{fig:ignite_13000_hard_3tenants} show that the two techniques meet the hard requirement for the tenants. 
However, the total cache size is {\em not} sufficient to meet the soft requirement.
In fair-sharing shown in Fig.~\ref{fig:fair_13000_ignite_hard_3tenants}, $u_1$ and $u_2$ manage their hit rates successfully for the soft requirement.
As soon as $u_3$ comes in, the sharing results in a noticeable drop of hit rate for $u_2$, since $u_2$ utilized a greater number of slots in $SC$ than $u_1$ (because $u_2$'s access pattern is less skewed than $u_1$'s).  
At the time $u_3$ leaves, the hit rates for $u_1$ and $u_2$ are restored and meet the defined soft requirement.
With selfish-sharing, even $u_2$ does not meet the soft requirement initially, since $u_1$ does not actively share in order to manage its hit rate to be over 70\%.
The sharing of $SC$ is somewhat limited over time as can be seen from Fig.~\ref{fig:selfish_13000_ignite_hard_3tenants}. 

%% file: conc.tex
\section{Conclusions}
	\label{sec:conc}

In this paper, we presented a new cache model that incorporates global caching and static caching to meet the per-tenant data access performance requirement with the optimized cache space use for a private cloud.
A cache insertion algorithm was also presented to efficiently manage the cache space in the proposed cache architecture, with two options of fair-sharing and selfish-sharing that can be chosen depending on the sharing policy among tenants.
An extensive set of experiments have been conducted on a simulation model and on a real setting equipped with an in-memory cache system (Apache Ignite).
The experimental results showed that the proposed cache model guarantees the defined performance requirement as long as the cache resources are sufficient to meet the requirement, although the existing techniques often fail to meet under the same condition.
Moreover, the proposed model requires much smaller cache space than the existing techniques, saving the cache space up to 49\% to satisfy the defined performance goal.

There are several interesting research topics for the future investigation.
An important problem is how to determine the size of the dedicated cache space.
A simple option is to refer to the required number of slots to meet the hard requirement under the assumption of the least skewed distribution for the entire users, but it can be further optimized using a prediction function as the dynamic caching performs. 
However, the prediction would be simpler than dynamic caching since it is only needed to determine the {\em minimal} cache size rather than the {\em optimal} cache size.
In addition, it is assumed a fixed length of cache slots to simplify the problem, but multiple slab classes may need to be considered for different types of applications requiring various slot sizes.
Another interesting topic would be the optimization for duplicated cache entries among tenants to better utilize the shared cache space.